\documentstyle[11pt,amssymb,epsfig,amsmath]{article}

\begin{document}
\newcommand{\be}{\begin{equation}}
\newcommand{\ee}{\end{equation}}
\newcommand{\ba}{\begin{eqnarray}}
\newcommand{\ea}{\end{eqnarray}}
\newcommand{\no}{\nonumber \\}
\newcommand{\gsim}{\mathrel{\hbox{\rlap{\lower.55ex \hbox {$\sim$}}
                   \kern-.3em \raise.4ex \hbox{$>$}}}}
\newcommand{\lsim}{\mathrel{\hbox{\rlap{\lower.55ex \hbox {$\sim$}}
                   \kern-.3em \raise.4ex \hbox{$<$}}}}

\def\be{\begin{eqnarray}}
\def\ee{\end{eqnarray}}
\def\bea{\be}
\def\eea{\ee}
\newcommand{\e}{{\mbox{e}}}
\def\del{{\partial}}
\def\vr{{\vec r}}
\def\vk{{\vec k}}
\def\vq{{\vec q}}
\def\vP{{\vec P}}
\def\vt{{\vec \tau}}
\def\vs{{\vec \sigma}}
\def\vJ{{\vec J}}
\def\vB{{\vec B}}
\def\hatr{{\hat r}}
\def\hatk{{\hat k}}
\def\roughly#1{\mathrel{\raise.3ex\hbox{$#1$\kern-.75em%
\lower1ex\hbox{$\sim$}}}}
\def\lsim{\roughly<}
\def\gsim{\roughly>}
\def\fm{{\mbox{fm}}}
\def\vx{{\vec x}}
\def\vy{{\vec y}}
\def\({\left(}
\def\){\right)}
\def\[{\left[}
\def\]{\right]}
\def\EM{{\rm EM}}
\def\barp{{\bar p}}
\def\zz{{z \bar z}}
\def\mus{{\cal M}_s}
\def\abs#1{{\left| #1 \right|}}
\def\ve{{\vec \epsilon}}
\def\nlo#1{{\mbox{N$^{#1}$LO}}}
\def\MS{{\mbox{M1V}}}
\def\mut{{\mbox{M1S}}}
\def\Qt{{\mbox{E2S}}}
\def\rM{{\cal R}_{\rm M1}}\def\rE{{\cal R}_{\rm E2}}
\def\la{{\Big<}}
\def\ra{{\Big>}}
\def\lsim{\mathrel{\rlap{\lower3pt\hbox{\hskip1pt$\sim$}}
     \raise1pt\hbox{$<$}}} 
\def\gsim{\mathrel{\rlap{\lower3pt\hbox{\hskip1pt$\sim$}}
     \raise1pt\hbox{$>$}}} 
\def\N{${\cal N}\,\,$}

\def\ka{\kappa}
\def\lam{\lambda}
\def\Lam{\Lambda}
\def\dlt{\delta}
\def\eps{\epsilon}
\def\sig{\sigma}
\def\omg{\omega}
\def\Omg{\Omega}
\def\vp{{\vec \xi}}
\def\vu{{\vec u}}
\def\avp{\overleftarrow \xi}
\def\avu{\overleftarrow u}
\def\pdu{\overleftarrow{\xi}\nabla\overleftarrow{u}}
\def\dvu{\dot{\vec u}}
\def\dvp{\dot{\vec \xi}}
\def\ve{{\vec E}}
\def\vb{{\vec B}}
\def\crs{\times}
\def\lv{\lvert}
\def\rv{\rvert}
\def\bsig{{\bar\sigma}}
\def\blam{{\bar\lambda}}

\def\J#1#2#3#4{ {#1} {\bf #2} (#4) {#3}. }
\def\PRL{Phys. Rev. Lett.}
\def\PL{Phys. Lett.}
\def\PLB{Phys. Lett. B}
\def\NP{Nucl. Phys.}
\def\NPA{Nucl. Phys. A}
\def\NPB{Nucl. Phys. B}
\def\PR{Phys. Rev.}
\def\PRC{Phys. Rev. C}

\renewcommand{\thefootnote}{\arabic{footnote}}
\setcounter{footnote}{0}

MPP-2011-49

\vskip 0.4cm \hfill { }
 \hfill {\today} \vskip 1cm

\begin{center}
{\LARGE\bf On the anomalous superfluid hydrodynamics
   }
\date{\today}

\vskip 1cm {\large Shu Lin\footnote{E-mail: slin@mppmu.mpg.de}
\\Max-Planck-Institut f\"{u}r Physik (Werner-Heisenberg-Institut)
\\ F\"{o}hringer Ring 6, 80805 M\"{u}nchen, Germany}


\end{center}

\vskip 0.5cm

\begin{center}

\end{center}

\vskip 0.5cm

\begin{abstract}
It has been shown by Son and Sur\'owka that the presence of anomaly in hydrodynamics
with global $U(1)$ symmetry can induce vortical and magnetic currents. The
induced current is uniquely determined by anomaly from
the existence of an entropy current with non-negative divergence. In this 
work, we extended the analysis to hydrodynamics with $U(1)$ symmetry 
spontaneously broken, i.e. $U(1)$ superfluid hydrodynamics. We found that
all possible first order gradient corrections are determined up to five
arbitrary functions, with the entropy current containing one arbitrary function. Furthermore, the stress tensor does not receive correction from terms proportional to the magnetic field.
\end{abstract}

\newpage

\renewcommand{\thefootnote}{\#\arabic{footnote}}
\setcounter{footnote}{0}

\section{Introduction}

The hydrodynamics is believed to be a universal description of quantum
field theory in long time and large distance limit. The assumption of
local thermal equilibrium allows a description of the
system in terms of fluid velocity $u^\mu(x)$ and 
local thermodynamical quantities including temperature
$T(x)$ and chemical potential $\mu(x)$ for conserved quantities, which
has the slowest relaxation. If the underlying quantum system has a
spontaneously broken symmetry, the resulting gapless Goldstone mode
should also be taken into account as an additional degree of freedom.
Normal hydrodynamics should be replaced by superfluid hydrodynamics,
the dynamics of which includes a superfluid component with velocity $\xi^\mu(x)$.
There have been significant efforts in formulating superfluid hydrodynamical 
description of QCD with spontaneous chiral symmetry breaking\cite{nuclear,zhang,lyon}. 
Superfluidity is also believed to be relevant for the phenomenology of
heavy ion collisions\cite{zakharov}.

When the system under consideration is not evolving slowly or smoothly enough,
the ideal hydrodynamical equations should be improved by a systematic inclusion of
of gradient terms, compatible with the symmetry of the system. The coefficients
appearing in front of the gradient terms are referred to as transport 
coefficients. For normal hydrodynamics, the transport coefficients corresponding
to first order gradient terms are
shear viscosity, bulk viscosity and charge diffusion constant. They should be
calculated from the microscopic theory. For example, in deconfined phase of
QCD or QCD-like theories, the transport coefficients have been
extensively studied in both weak coupling\cite{AMY} and strong coupling 
regime\cite{PSS,rcharge,buchel,kharzeev}. For superfluid hydrodynamics, the
number of transport coefficients increases significantly due to the additional
superfluid velocity. It has been recently enumerated in \cite{Minwalla}, 
see also \cite{HY} that there are $14$ 
transport coefficients for $U(1)$ superfluid.
The transport coefficients have also been calculated by method of
gauge/gravity duality.

While previous studies have been focusing on the parity even term in the
gradient expansion, parity odd terms can be equally important.
Early studies of fluid/gravity duality found an unexpected term 
proportional to the vorticity of the fluid in the constitutive equation
of the R-current\cite{haack,bbbdls}. The puzzling term seems to lead to
negative entropy production at the first sight. This issue is clarified by
Son and Sur\'owka in \cite{SS}, where they showed 
that the vortical term, as well as a
term proportional to the magnetic field is actually required by
the existence of entropy current with non-negative divergence. 
Working to the first order in
gradient expansion, they found that in the presence of global $U(1)$ anomaly,
both the charge current and the entropy current get modified by a vortical
and a magnetic term. Remarkably, the form of the correction is almost
entirely determined by the anomaly. 
This result was generalized to nonabelian symmetry group in \cite{oz},
where similar effect has been found. 

In this work, we would like to consider the effect of triangle anomaly on
the $U(1)$ superfluid hydrodynamics. We will focus on the parity odd sector
and use the existence of entropy current to determine the 
transport coefficients. We will first write down the ideal superfluid
hydrodynamical equations in external $U(1)$ field in Section 2.
We will construct all possible first gradient terms in parity odd sector
in Section 3, following the technique in \cite{Minwalla}. The constraint
equations on the transport coefficients will be derived based on the
criterion of non-negative entropy production in Section 4. The
constraint equations are solved in Section 5 and we will conclude in Section 6.

\section{Ideal superfluid hydrodynamics in the presence of external field}

The nonrelativistic hydrodynamical equation for $U(1)$ superfluid in
the absence of external field is 
known as Landau's two fluid model\cite{landau}. It can be derived with different
approaches, including quasi-average\cite{bogoliubov}, Poisson bracket\cite{volovick}
and effective Lagrangian\cite{GWW,peletminskii}.
The generalization to relativistic superfluid hydrodynamics
is written down in \cite{Carter,LK,CK,Israel,son_u1}.
We quote the result of ideal relativistic superfluid hydrodynamics of 
\cite{son_u1} in the following:

\begin{align}\label{u1_noA}
\left\{\begin{array}{l}
\del_\mu T^{\mu\nu}=0 \\
\del_\mu j^\mu=0 \\
\mu+u^\mu\del_\mu\varphi=0.
\end{array}
\right.
\end{align}

The first and second equation in (\ref{u1_noA}) is energy-momentum conservation
and current conservation respectively. The third equation is a ``Josephson''
equation. It is a statement that the density of the fluid and phase of the
condensate is not independent, with $\varphi$ being the phase of the condensate.
The constitutive equations are:

\begin{align}
&T^{\mu\nu}=(\eps+p)u^\mu u^\nu+p\eta^{\mu\nu}+f^2\del^\mu\varphi\del^\nu\varphi \\
&j^\mu=nu^\mu+f^2\del^\mu\varphi ,
\end{align}

together with the equation of state: $d\eps=Tds+\mu dn+f^2dX$ and
$dp=sdT+nd\mu-f^2dX$. Here $X=\frac{1}{2}(\del\varphi)^2$ is the additional thermodynamical
variable due to the superfluidity. The entropy is conserved in the
ideal superfluid hydrodynamics:

\begin{align}
\del_\mu(su^\mu)=0.
\end{align}

In the presence of an external $U(1)$ field(provided that
it does not destroy the superfluidity), the hydrodynamical 
equations need to be modified. As in normal fluid, external 
electromagnetic field will exert Coulomb and
Lorentz force on the
fluid. The energy-momentum is not conserved as in the isolate system of
superfluid. A term of
 $F^{\nu\lam}j_\lam$ is needed on the RHS of (\ref{u1_noA})
to account for it. Furthermore,
assuming the external $U(1)$ field couples to the
matter field through minimal coupling in the microscopic theory,
the superfluid velocity receives contribution from the external field
in the following way:
$\nabla\varphi\rightarrow\nabla\varphi-\vec{A}$. Taking into
account relativistic invariance, it is not difficult to convince
oneself that the hydrodynamical equations of superfluid are given by:

\begin{align}\label{u1_A}
\left\{\begin{array}{l}
\del_\mu T^{\mu\nu}=F^{\nu\lam}j_\lam \\
\del_\mu j^\mu=0 \\
\mu=-\xi\cdot u \\
\del_\mu \xi_\nu-\del_\nu \xi_\mu=-F_{\mu\nu}
\end{array}
\right.,
\end{align}

with $T^{\mu\nu}=(\eps+p)u^\mu u^\nu+p\eta^{\mu\nu}+f^2\xi^\mu \xi^\nu$ and
$j^\mu=nu^\mu+f^2\xi^\mu$. The equation of state is the same as the case without
external field: $d\eps=Tds+\mu dn+f^2dX$ and
$dp=sdT+nd\mu-f^2dX$, but with $X=\frac{1}{2}\xi^2$. 
The phase of the condensate
$\varphi$ is replaced by $\xi_\mu$, the un-normalized
superfluid velocity.
We have confirmed the intuitive equation (\ref{u1_A}) with Poisson bracket
method\cite{lin}.
Starting from ideal hydrodynamical equations, we can show that 
there is a conserved entropy current: The first two equations of (\ref{u1_A})
give,

\begin{align}\label{s_ideal}
\frac{1}{T}u_\nu\del_\mu T^{\mu\nu}+\frac{\mu}{T}\del_\mu j^\mu=\frac{1}{T}u_\nu F^{\nu\lam}j_\lam .
\end{align}

Upon using $\del_\mu \xi_\nu-\del_\nu \xi_\mu=-F_{\mu\nu}$ and the equation of
state, we readily obtain $\del_\mu(su^\mu)=0$.

\section{First order gradient expansion in the parity odd sector}

In this section, we want to study the
effect of axial anomaly on superfluid hydrodynamics, which amounts to
a modification to the current conservation equation: 
$\del_\mu j^\mu=CE\cdot B$, with $E^\mu=F^{\mu\nu}u_\nu$ and 
$B^\mu=\frac{1}{2}\eps^{\mu\nu\alpha\beta}u_\nu F_{\alpha\beta}$.
\footnote{There is also an interesting effect due to gravitational anomaly, 
see \cite{landsteiner}. We do not discuss it here}. In this work, we will
restrict ourselves to a single $U(1)$ charge. The anomaly can be realized as
an AAA anomaly in a fictitious $U(1)$ theory: 
${\cal L}={\bar \psi}\gamma^\mu(\del_\mu+ie\gamma^5 A_\mu)\psi$, with $\psi$ and
$A_\mu$ being the matter field and external axial $U(1)$ field respectively.
It will be interesting to investigate the case of two charges for phenomelogical
applications \cite{KS,CMW,stu,russian}. We leave it for future work.
Note that the superfluidity
arises from spontaneous symmetry breaking. We should ensure that the
$U(1)$ symmetry
(charge conservation) is not spoiled by the anomaly term $CE\cdot B$
\footnote{We thank Ingo Kirsch for pointing this out to us}. We can
for example turn off the electric field, keeping only the magnetic field, so
that the anomaly term vanishes. However its effect on charge current and
entropy current remains, as is clear from \cite{SS}.

The first order gradient expansion will introduce correction to the 
stress tensor, current and chemical potential\cite{HY,Minwalla}:

\begin{align}
&T^{\mu\nu}=(\eps+p)u^\mu u^\nu+p\eta^{\mu\nu}+f^2\xi^\mu \xi^\nu+\pi^{\mu\nu} \\
&j^\mu=nu^\mu+f^2\xi^\mu+\nu^\mu \\
&\mu=-\xi\cdot u+\mu_A .
\end{align}

We can specify a frame to further constrain the corrections. The frames 
often used are the fluid frame, where $\pi_{\mu\nu}u^\mu=0$ and $\mu_A=0$ and the
transverse frame, where $\pi_{\mu\nu}u^\mu=0$ and $\nu_\mu u^\mu=0$\cite{Minwalla}. 
We will choose to work in the transverse frame in the following.

The next step is to to write down all possible parity odd, 
 first order gradient terms 
that can appear in $\pi_{\mu\nu}$, $\nu_\mu$ and $\mu_A$. These include gradient 
of the fluid velocity($\del u_\mu$ and $\del \xi_\mu$) and gradient of the 
thermodynamical variables($\del p$, $\del\mu$ and $\del X$). However it is
complicated by the fact that they are not completely independent, but
related by ideal hydrodynamical equations. The procedure we will use in
 enumerating independent gradient terms closely
follow the method adopted in \cite{Minwalla}: We first list all possible
(pseudo)scalars, (pseudo)vectors and (pseudo)tensors,
which are independent
upon using the ideal hydrodynamical equations. Then we construct $\pi^{\mu\nu}$,
$\nu_\mu$ and $\mu_A$ out of them.

Assuming the external $U(1)$ field $A_\mu\sim O(p^0)$, the 
field strength $F_{\mu\nu}=\del_\mu A_\nu-\del_\nu A_\mu$ is first order in gradient.
There are in total $2$ pseoduscalars, $7$ pseudovectors, $2$ pseudotensors,
$7$ scalars, $7$ vectors and $2$ tensors that are first order in gradient. 
Note that due to the external field
the number of independent structures are slightly larger 
than those reported in \cite{Minwalla}, where
the numbers are $1,\,5,\,6,\,5$ for pseudoscalar, pseudovector, scalar and
vector respectively.
The independent structures are listed as follows.

pseudotensor:
\begin{align}
&\bsig^{\mu\nu}=\eps^{\mu\rho\alpha\beta}u_\rho\xi_\alpha\sig_\beta^\nu+\eps^{\nu\rho\alpha\beta}u_\rho\xi_\alpha\sig_\beta^\mu \no
&\blam^{\mu\nu}=\eps^{\mu\rho\alpha\beta}u_\rho\xi_\alpha\lam_\beta^\nu+\eps^{\nu\rho\alpha\beta}u_\rho\xi_\alpha\lam_\beta^\mu
\end{align}

pseudovector:
\begin{align}
&\omg^\mu=\frac{1}{2}\eps^{\mu\nu\alpha\beta}u_\nu\del_\alpha u_\beta,\;
B^\mu\no
&\Omg^\mu=\frac{1}{2}P^{\mu\sig}\eps_{\sig\nu\alpha\beta}\xi^\nu\del^\alpha u^\beta-\mu\omg^\mu,\;
K^\mu=\frac{1}{2}P^{\mu\sig}\eps_{\sig\nu\alpha\beta}\xi^\nu F^{\alpha\beta}-\mu B^\mu\no
&U^\mu=\frac{1}{2}\eps^{\mu\nu\alpha\beta}u_\nu \xi_\alpha\del_\beta p,\;
V^\mu=\frac{1}{2}\eps^{\mu\nu\alpha\beta}u_\nu \xi_\alpha\del_\beta \mu\no
&L^\mu=\frac{1}{2}\xi^\lam\eps^{\mu\nu\alpha\beta}u_\nu \xi_\alpha\del_\beta u_\lam,
\end{align}

pseudoscalar:
\begin{align}
\xi\cdot\omg,\;\xi\cdot B,
\end{align}

tensor:
\begin{align}
&\sig_{\mu\nu}=P^\alpha_\mu P^\beta_\nu\(\frac{\del_\alpha u_\beta+\del_\beta u_\alpha}{2}-\eta_{\alpha\beta}\frac{\del\cdot u}{3}\) \no
&\lam_{\mu\nu}=P^\alpha_\mu P^\beta_\nu\(\frac{\del_\alpha \xi_\beta+\del_\beta \xi_\alpha}{2}-\eta_{\alpha\beta}\frac{\del\cdot \xi}{3}\)
\end{align}

vector:
\begin{align}
\sig_{\mu\nu}l^\nu,\;\del_\mu p,\;\del_\mu\mu,\;\del_\mu X,\;\xi_\nu\del_\mu u^\nu,\;E_\mu,\;F_{\mu\nu}\xi^\nu
\end{align}

scalar:
\begin{align}
&\zeta=\del_\mu(nu^\mu),\,\theta=\del_\mu u^\mu,\,\sig_{\mu\nu}l^\mu l^\nu,\no
&l^\mu\del_\mu p,\, l^\mu\del_\mu X,\, u^\mu\del_\mu X,\,E\cdot \xi,
\end{align}
where $P^{\mu\nu}=\eta^{\mu\nu}+u^\mu u^\nu$ is the projection operator,
which is transverse to the normal fluid velocity $u_\mu P^{\mu\nu}=0$.
$l^\mu=P^{\mu\nu}\xi_\nu$ is the component of $\xi_\mu$
transverse to the normal fluid velocity.

Now we are ready to decompose $\pi^{\mu\nu}$, $\nu^\mu$ and $\mu_A$ in terms
of the independent structures. Note the constraint 
$u_\mu\pi^{\mu\nu}=0,\,u_\mu\nu^\mu=0$ from choosing the transverse frame, we
have:

\begin{align}\label{para}
&\pi^{\mu\nu}=T_\sig\bsig^{\mu\nu}+T_\lam\blam^{\mu\nu}+P^{\mu\nu}(P_\omg \xi\cdot\omg+P_B \xi\cdot B)+2(\Pi_\omg l^{(\mu}\omg^{\nu)}+\Pi_B l^{(\mu}B^{\nu)}+\Pi_\Omg l^{(\mu}\Omg^{\nu)} \no
&+\Pi_K l^{(\mu}K^{\nu)}+\Pi_U l^{(\mu}U^{\nu)}+\Pi_V l^{(\mu}V^{\nu)}+\Pi_L l^{(\mu}L^{\nu)})+l^\mu l^\nu(N_\omg \xi\cdot\omg+N_B \xi\cdot B) \\
&\nu^\mu=(V_\omg\omg^\mu+V_B B^\mu+V_\Omg \Omg^\mu+V_K K^\mu+V_U U^\mu+V_V V^\mu+V_L L^\mu) 
+(\Lam_\omg \xi\cdot\omg+\Lam_B \xi\cdot B)l^\mu \\
&\mu_A=A_\omg \xi\cdot\omg+A_B \xi\cdot B,
\end{align}
where $A^{(\mu}B^{\nu)}=\frac{1}{2}\(A^\mu B^\nu+A^\nu B^\mu\)$.

\section{Constraint from entropy conservation}

In the presence of triangle anomaly, the charge current is no longer conserved:
$\del_\mu j^\mu=CE\cdot B$. This enters as a second order correction in
gradient to the hydrodynamical equations, thus we need to use the 
constitutive equations up to first order in gradient expansion, which has
been worked out in the last section. Therefore, we have:

\begin{align}
\left\{\begin{array}{l}
\del_\mu T^{\mu\nu}=F^{\nu\lam}j_\lam \\
\del_\mu j^\mu=CE\cdot B \\
\mu+\mu_A=-\xi\cdot u \\
\del_\mu \xi_\nu-\del_\nu \xi_\mu=-F_{\mu\nu}
\end{array}
\right.,
\end{align}
Note an explicit change to the form of ``Josephson'' equation due to the
gradient correction. This is because we will choose $\mu$ as a separate 
thermodynamical variable, which is always zeroth order in gradient.
While the gradient corrections to the energy-momentum and current 
conservation are implicit through the constitutive equations:
$T^{\mu\nu}=(\eps+p)u^\mu u^\nu+p\eta^{\mu\nu}+f^2\xi^\mu\xi^\nu+\pi^{\mu\nu}$ and
$j^\mu=nu^\mu+f^2 \xi^\mu+\nu^\mu$. In the presence of first order gradient
correction and anomaly, the entropy current defined in (\ref{s_ideal}) is
not conserved anymore. It satisfies the following equation:

\begin{align}\label{s_div}
\del_\mu(su^\mu-\frac{\mu}{T}\nu^\mu)=-\frac{1}{T}\del_\mu u_\nu\pi^{\mu\nu}
-\del_\mu\(\frac{\mu}{T}\)\nu^\mu+\frac{\mu_A}{T}\del_\mu(nu^\mu)+\frac{E\cdot\nu}{T}-\frac{\mu}{T}CE\cdot B.
\end{align}

We should modify the definition of the entropy current and 
the positivity of entropy production will allow us to fix the transport
coefficients. The parity even sector has been investigated in \cite{Minwalla,HY}. In particular, this criterion can significantly reduce the number of nonvanishing
transport coefficients. We are interested in the parity odd sector. Remarkably,
in case of normal hydrodynamics, Son and Sur\'owka showed the positivity
(rather vanishing for the parity odd sector) allows them to almost 
completely fix the transport coefficients. As we will see soon, this is
also true for superfluid hydrodynamics.

We start by writing down the most general form of the entropy current, up
to first order in gradient:
\begin{align}\label{s_vis}
s^\mu&=su^\mu-\frac{\mu}{T}\nu^\mu+s_\omg\omg^\mu+s_B B^\mu+s_\Omg \Omg^\mu+s_K K^\mu+s_U U^\mu+s_V V^\mu+s_L L^\mu \no
&+(s_{n\omg}\xi\cdot\omg+s_{nB}\xi\cdot B)l^\mu+(s_{u\omg}\xi\cdot\omg+s_{uB}\xi\cdot B)u^\mu.
\end{align}
The corresponding entropy production is given by:
\begin{align}\label{s_vis_ex}
&\del_\mu s^\mu=-\frac{1}{T}\del_\mu u_\nu\pi^{\mu\nu}
-\del_\mu\(\frac{\mu}{T}\)\nu^\mu+\frac{\mu_A}{T}\del_\mu(nu^\mu)\zeta+\frac{E\cdot\nu}{T}-\frac{\mu}{T}CE\cdot B +\del_\mu(s_\omg\omg^\mu+s_B B^\mu+ \no
&s_\Omg \Omg^\mu+s_K K^\mu+s_U U^\mu+s_V V^\mu+s_L L^\mu
+(s_{n\omg}\xi\cdot\omg+s_{nB}\xi\cdot B)l^\mu+(s_{u\omg}\xi\cdot\omg+s_{uB}\xi\cdot B)u^\mu).
\end{align}

We first note that all terms in the expansion of (\ref{s_vis_ex}) are
parity odd, second order in gradient. Most of them can be decomposed on
a basis in the form: pseudoscalar$\cdot$scalar, pseudovector$\cdot$vector,
and pseudotensor$\cdot$tensor, with the independent structures listed in the
previous section. However, the divergences of the last four terms of the 
entropy current contain new structures with double derivatives: 
$l^\mu\xi^\nu\del_\mu\omg_\nu$, $l^\mu\xi^\nu\del_\mu B_\nu$, $u^\mu\xi^\nu\del_\mu\omg_\nu$
 and $u^\mu\xi^\nu\del_\mu B_\nu$. Only three of the four are independent of 
each other. The last two are related by ideal hydrodynamical equations. We
start with the following identity from ideal hydrodynamical equations:
\begin{align}
\del_\mu((\eps+p)u^\mu u_\nu)+\del_\nu p+f^2\del_\nu X-\xi_\nu\del_\mu(nu^\mu)=nE_\nu,
\end{align}
where $w=\eps+p$ is the enthalpy. Applying $\eps^{\nu\lam\alpha\beta}\xi_\alpha u_\beta\del_\lam$ to the above, we obtain:
\begin{align}\label{dd}
&2\xi\cdot\omg\del_\mu(wu^\mu)-2\Omg^\mu\del_\mu w+2w(\xi\cdot\omg\theta-\omg^\mu l^\nu\del_\mu u_\nu)+w\eps^{\mu\nu\alpha\beta}\xi_\mu u_\nu u^\rho\del_\rho\del_\alpha u_\beta +\xi\cdot B\del_\mu(nu^\mu) \no
&-\eps^{\mu\nu\alpha\beta}\del_\mu f^2u_\nu\xi_\alpha\del_\beta X-K^\mu\del_\mu n+n\eps^{\mu\nu\alpha\beta}\xi_\mu u_\nu u^\rho\del_\rho\del_\alpha A_\beta 
-n(B^\mu l^\nu\del_\mu u_\nu-\xi\cdot B\theta)=0.
\end{align}
In addition, $s_\Omg\del_\mu\Omg^\mu$ and $s_K\del_\mu K^\mu$ also contain the 
following contributions to the structure $u^\mu\xi^\nu\del_\mu\omg_\nu$
 and $u^\mu\xi^\nu\del_\mu B_\nu$:
\begin{align}\label{sok}
&-\frac{1}{2}s_\Omg\eps^{\mu\nu\alpha\beta}\xi_\mu u_\nu u^\rho\del_\rho\del_\alpha u_\beta \no
&-s_K\eps^{\mu\nu\alpha\beta}\xi_\mu u_\nu u^\rho\del_\rho\del_\alpha A_\beta \nonumber.
\end{align}
The requirement of the vanishing entropy production on the three independent
basis forces: $s_{n\omg}=s_{nB}=0$, while the other two entropy current 
coefficients are related by:
\begin{align}
s_{u\omg}=2ws_u+s_\Omg\quad s_{uB}=ns_u+s_K.
\end{align}

We can then proceed to decompose all the terms on the
basis given by pseudotensor $\cdot$tensor, pseudoscalar$\cdot$scalar and 
pseudovector$\cdot$vector. In achieving this goal, we will need 
the following identities:
\begin{align}
&\del_\mu u_\nu\pi^{\mu\nu}=\sig_{\mu\nu}\pi^{\mu\nu}+\frac{\del\cdot u}{3}\pi^{\theta}_\theta\\
&\del_\mu\(\frac{\mu}{T}\)=\frac{\del(\mu/T)}{\del p}\del_\mu p+\frac{\del(\mu/T)}{\del \mu}\del_\mu \mu+\frac{\del(\mu/T)}{\del X}\del_\mu X.
\end{align}
It is easy to obtain from $dp=sdT+nd\mu-f^2dX$ that
\begin{align}
\frac{\del(\mu/T)}{\del p}=-\frac{\mu}{sT^2}\quad\frac{\del(\mu/T)}{\del \mu}=\frac{w}{sT^2}\quad\frac{\del(\mu/T)}{\del X}=-\frac{\mu}{sT^2}f^2.
\end{align}
The last term $\del_\mu(s_OO^\mu)$($O=\omg,B,\Omg,K,U,V,L$) generates 
$O^\mu\del_\mu(s_O)$ and $s_O\del_\mu O^\mu$. The former can be brought into
the form vector$\cdot$pseudovector via

\begin{align}
\del_\mu(s_O)=\frac{\del s_O}{\del p}\del_\mu p+\frac{\del s_O}{\del \mu}\del_\mu \mu+\frac{\del s_O}{\del X}\del_\mu X,
\end{align}
while the latter needs some extra work. By ideal hydrodynamical equations,
$\del_\mu O^\mu$ can be expressed as a linear combination of 
pseudoscalar$\cdot$scalar, pseudovector$\cdot$vector and 
pseudotensor $\cdot$tensor. The only exceptions are $\del_\mu\Omg^\mu$ and 
$\del_\mu K^\mu$, which contain terms with double derivatives and are used to fix
the entropy current coefficients above. We listed the
final results below and move the details to the appendix.

\begin{align}
&\del_\mu\omg^\mu=\frac{2}{w}(nE\cdot\omg-\omg^\mu\del_\mu p-f^2\omg^\mu\del_\mu X+\xi\cdot\omg\zeta) \\
&\del_\mu B^\mu=\frac{1}{w}(nE\cdot B-B^\mu\del_\mu p-f^2B^\mu\del_\mu X+\xi\cdot B\zeta)-2E\cdot\omg \\
&\del_\mu\Omg^\mu=-\theta \xi\cdot\omg-\frac{1}{2w}(-B^\mu\del_\mu p-f^2B^\mu\del_\mu X+nE\cdot B+\xi\cdot B\zeta)-\frac{\mu}{w}(nE\cdot\omg+\xi\cdot\omg\zeta \no
&-\omg^\mu\del_\mu p-f^2\omg^\mu\del_\mu X)+\omg^\mu \xi^\nu\del_\mu u_\nu
-\frac{1}{2}\eps^{\mu\nu\alpha\beta}\xi_\mu u_\nu u^\rho\del_\rho\del_\alpha u_\beta \\
&\del_\mu K^\mu=-\theta \xi\cdot B+2\mu E\cdot\omg+\frac{1}{w}(-K^\mu\del_\mu p-f^2K^\mu\del_\mu X)+\frac{3}{2}E\cdot B-B^\mu\del_\mu \mu \no
&-\eps^{\mu\nu\alpha\beta}\xi_\mu u_\nu u^\rho\del_\rho\del_\alpha A_\beta\\
&\del_\mu U^\mu=\frac{1}{2w}(nK^\mu\del_\mu p-2f^2U^\mu\del_\mu X)-\mu\omg^\mu\del_\mu p-\xi\cdot\omg u^\nu\del_\nu p-\frac{1}{2}B^\mu\del_\mu p\\
&\del_\mu V^\mu=\frac{1}{2w}(nK^\mu\del_\mu\mu+2U^\mu\del_\mu\mu-2f^2V^\mu\del_\mu X)-\mu\omg^\mu\del_\mu\mu-\xi\cdot\omg u^\mu\del_\mu\mu-\frac{1}{2}B^\mu\del_\mu\mu\\
&\del_\mu L^\mu=\frac{1}{w}(\frac{nK^\mu \xi^\nu\del_\mu u_\nu}{2}-L^\mu\del_\mu p-f^2L^\mu\del_\mu X)-\mu\omg^\mu \xi^\nu\del_\mu u_\nu -\frac{\xi\cdot\omg}{w}(nE\cdot \xi-l^\mu\del_\mu p \no
&-f^2l^\mu\del_\mu X+l^2\zeta)-\frac{1}{2}B^\mu \xi^\nu\del_\mu u_\nu+\frac{1}{2}\del_\mu\xi^\lam\eps^{\mu\nu\alpha\beta}u_\nu\xi^\alpha\del_\beta u_\lam.
\label{divergence}
\end{align}
\newline
Note in the above, $u^\mu\del_\mu p$, $u^\mu\del_\mu\mu$ are not listed as
independent scalars in the previous section. They can be expressed as:
\begin{align}
&u^\mu\del_\mu \mu=b_s(-\theta s)+b_n(\zeta-n\theta)+b_X u^\mu\del_\mu X\\
&u^\mu\del_\mu p=c_s(-\theta s)+c_n(\zeta-n\theta)+c_X u^\mu\del_\mu X,
\end{align}
where $b_s=\(\frac{\del\mu}{\del s}\)_{n,X}$, $b_n=\(\frac{\del\mu}{\del n}\)_{s,X}$
, $b_X=\(\frac{\del\mu}{\del X}\)_{s,n}$ and $c_s=\(\frac{\del p}{\del s}\)_{n,X}$,
 $c_n=\(\frac{\del p}{\del n}\)_{s,X}$, $c_X=\(\frac{\del p}{\del X}\)_{s,n}$.

Let us consider the basis of the form pseudotensor$\cdot$tensor first. The only
basis get popularized are $\sig_{\mu\nu}\bsig^{\mu\nu}$ and $\sig_{\mu\nu}\blam^{\mu\nu}$.
The former receives the contribution from $T_\sig\sig_{\mu\nu}\bsig^{\mu\nu}$, which 
vanished identically thus does not constrain $T_\sig$\cite{BBMY}. On the
other hand, the latter receives contributions from $T_\lam\sig_{\mu\nu}\blam^{\mu\nu}$
and $s_L\del_\mu L^\mu$. The last term of $\del_\mu L^\mu$ as shown in 
(\ref{divergence}) should be expressed as follows:

\begin{align}
&\frac{1}{2}\del_\mu\xi^\lam\eps^{\mu\nu\alpha\beta}u_\nu\xi^\alpha\del_\beta u_\lam=-\frac{1}{2}\bigg[\frac{1}{2}\sig_{\mu\nu}\blam^{\mu\nu}+\xi\cdot\omg (\frac{\zeta+l^\mu\del_\mu f^2+\mu u^\mu\del_\mu f^2}{f^2}+u^\mu\del_\mu\mu+ \no
&\frac{nE\cdot\xi-l^\mu\del_\mu p-f^2l^\mu\del_\mu X+l^2\zeta}{w})-\frac{1}{3}\xi\cdot B\theta+\frac{1}{2}\sig_{\mu\nu}l^\mu B^\nu+\omg^\mu(\del_\mu X+\mu(\del_\mu\mu+\xi^\nu\del_\mu u_\nu))\bigg].
\end{align}
The vanishing of the component of entropy production on the basis 
$\sig_{\mu\nu}\blam^{\mu\nu}$ gives:
\begin{align}
T_\lam=-\frac{T}{4}s_L.
\end{align}

Now we are ready to move to the basis of the form pseudoscalar$\cdot$scalar and
pseudovector $\cdot$vector. There is a last subtlety that not all 
pseudovector$\cdot$vector 
are linearly independent. Some of them can be expressed as a linear combination
of others and possibly also pseudoscalar $\cdot$scalar . We work out
the relevant relations in the appendix. The complete basis set we will contains
$2\times7$ pseudoscalar$\cdot$
scalar and $21$ pseudovector$\cdot$vector. The latter is listed as follows:
\begin{align}\label{basis}
&\sig_{\mu\nu}l^\mu B^\nu,\,\omg^\mu \xi^\nu\del_\mu u_\nu,\,B^\mu \xi^\nu\del_\mu u_\nu,\,K^\mu \xi^\nu\del_\mu u_\nu,\,\omg^\mu l^\nu\del_\mu p,\,B^\mu l^\nu\del_\mu p,\,K^\mu l^\nu\del_\mu p,\,L^\mu l^\nu\del_\mu p, \no
&\omg^\mu l^\nu\del_\mu\mu,\,B^\mu l^\nu\del_\mu\mu,\,K^\mu l^\nu\del_\mu\mu,\,L^\mu l^\nu\del_\mu\mu,\,U^\mu l^\nu\del_\mu\mu,\,\omg^\mu l^\nu\del_\mu X,\,B^\mu l^\nu\del_\mu X,\,K^\mu l^\nu\del_\mu X,\,L^\mu l^\nu\del_\mu X,\no
&U^\mu l^\nu\del_\mu X,\,V^\mu l^\nu\del_\mu X,\,E\cdot\omg,\,E\cdot B.
\end{align}

Plugging (\ref{divergence}) to (\ref{s_vis_ex}) and using (\ref{pvvpss})
-(\ref{pvvpsse}) to
organize all terms into the above basis, we obtain a set of $35$ coupled 
equations upon demanding the non-negativity of the entropy production(Clearly
all coefficients in the expansion on the basis have to vanish). We will
not spell out the full equations, but only comment on one property of the
equations. They are all linear in the unknowns, algebraic for the
transport coefficients and first order PDE for the entropy coefficients
$s_O$($O=\omg,B,\Omg,K,U,V,L,u$). This property will be crucial in solving the
equations.

\section{Solving the coupled equations}

As is noted in the previous section that
 the coupled equations are first order PDE for the entropy coefficients,
but are only algebraic for the transport coefficients. We can simplify the
problem by solving the transport coefficients in terms of the entropy
coefficients. This step is relatively straightforward. We end up with the
following results:

\begin{align}\label{reduced}
&s_L=0, \quad \Pi_B=0, \quad \Pi_U=0 \no
&\frac{l^2}{2}\Pi_L=\Pi_\omg,\quad \mu V_L=\frac{sT}{w}\Pi_\Omg \no
&\Pi_K=\frac{\Pi_V}{2},\quad \Pi_\Omg=\mu\Pi_V \no
&\frac{\mu V_\omg}{sT^2}-\frac{l^2\Pi_\Omg}{2Tw}=-\frac{\mu}{sT}\(2\mu s_K-2s_B+2sTs_u+\frac{2ns_\omg}{w}+(1-\frac{2n\mu}{w})s_\Omg\) \no
&V_B=-T\(-\frac{C\mu}{T}+(2-\frac{3n\mu}{2w})s_K+\frac{ns_B}{w}+\frac{nsT}{w}s_u-\frac{ns_\Omg}{2w}\) \no
&\frac{T(nb_n+sb_s)}{w(b_sc_n-b_nc_s)}s_u+\frac{ns_U}{2w}+\frac{\mu V_K}{sT^2}+\frac{V_U}{2T}-\frac{V_\Omg}{2sT^2}+\del_p s_K-\frac{n}{2w}\del_p s_\Omg=0 \no
&\(\frac{-2Tc_n+2\mu c_s}{w(b_sc_n-b_nc_s)}\)s_u+\frac{s_V}{w}-\frac{wV_U}{sT^2}-\frac{\mu V_V}{sT^2}+\frac{V_\Omg}{sT^2}-\del_ps_V+\del_\mu s_U-\frac{1}{w}\del_\mu s_\Omg=0 \no
&\frac{\mu V_U}{sT^2}+\(\frac{2\mu(1+\frac{c_x}{f^2})(nb_n+sb_s)-\frac{2\mu c_X}{f^2}(nb_n+sb_s)}{ws(b_sc_n-b_nc_s)}\)s_u+\(\frac{1}{f^2}\del_X-\frac{1}{w}\)s_U \no
&+\(\frac{1}{w}\del_p-\frac{1}{wf^2}\del_X\)s_\Omg=0 \no
&P_B=-nT\(1+(nc_n+sc_s)\del_p+(nb_n+sb_s)\del_\mu\)s_u-T\((nc_n+sc_s)\del_p+(nb_n+sb_s)\del_\mu\)s_K \no
&-\frac{w\mu(nb_n+sb_s)}{sT}\Lam_B \no
&N_B=\frac{w}{sT}\Lam_B \no
&\Lam_B=-\frac{sT^2}{w(1+\mu b_X)}\((\del_X+c_X\del_p+b_X\del_\mu)s_K+n(\del_X+c_X\del_p+b_X\del_\mu)s_u\) \no
&A_B=-T\bigg[-\frac{3\mu s_K}{2w}+\frac{s_B}{w}-\frac{n\mu}{w}s_u-\frac{s_\Omg}{2w}+\frac{\mu}{sT^2}(l^2+wb_n)\Lam_B+\(c_n\del_p+b_n\del_\mu\)s_K \no
&+n\(c_n\del_p+b_n\del_\mu\)s_u\bigg] \no
&\Lam_\omg+\frac{sT}{2w}\Pi_V=-\frac{sT^2}{w(1+\mu b_X)}\bigg[2w\(\del_X+c_X\del_p+b_X\del_\mu\)s_u+\(\del_X+c_X\del_p+b_X\del_\mu\)s_\Omg \no
&-c_Xs_U-b_Xs_V\bigg] \no
&N_\omg+\frac{\Pi_L}{2}=\frac{w}{sT}\(\Lam_\omg+\frac{sT}{2w}\Pi_V\) \no
&P_\omg=-2wTs_u+T(nc_n+sc_s)s_U+T(nb_n+sb_s)s_V-\frac{w\mu}{sT}(nb_n+sb_s)\(\Lam_\omg+\frac{sT}{2w}\Pi_V\) \no
&-2wT\((nc_n+sc_s)\del_p+(nb_n+sb_s)\del_\mu\)s_u-T\((nc_n+sc_s)\del_p+(nb_n+sb_s)\del_\mu\)s_\Omg \no
&A_\omg=-T\bigg[-2\mu s_u-c_ns_U-b_ns_V+\frac{2s_\omg}{w}-\frac{2\mu s_\Omg}{2}+\frac{\mu}{sT^2}(l^2+wb_n)(\Lam_\omg+\frac{sT}{2w}\Pi_V) \no
&+2w(c_n\del_p+b_n\del_\mu)s_u+(c_n\del_p+b_n\del_\mu)s_\Omg\bigg]
\end{align}

We find the transport coefficients are not fully determined by (\ref{reduced})
in terms of entropy coefficients $s_K,s_B,s_\omg,s_U,s_V,s_u$. In particular, we can
choose $\Pi_V$, $\Pi_\omg$, and $V_\Omg$ as the free inputs, in addition to $T_\sig$
found in the previous section. 

The entropy coefficients, independent of (\ref{reduced}), satisfy the following
equation:
\begin{subequations}
\begin{align}
&s_U=-\frac{1}{\mu}\(\frac{1}{f^2}\del_X-\del_p\)s_\omg \label{s1}\\
&s_U=-2\(\frac{1}{f^2}\del_X-\del_p\)s_\omg \label{s2}\\
&s_\Omg-\mu s_V=\(\frac{2}{\mu}-\frac{w}{\mu f^2}\del_X-\del_\mu\)s_\omg \label{s3}\\
&s_\Omg-\mu s_V=-\mu s_K+2\(1-\frac{w}{f^2}\del_X-\mu\del_\mu\)s_B \label{s4}\\
&s_\Omg=\(\frac{2}{\mu}-\frac{sT}{\mu f^2}\del_X\)s_\omg+2\mu s_K-2s_B \label{s5}\\
&s_\Omg=-\frac{2C\mu^2}{T}+\(2-\frac{2sT}{f^2}\del_X\)s_B+\mu s_K \label{s6}\\
&-\(\frac{1}{f^2}\del_X-\del_p\)s_K+\frac{1}{2\mu}\(\frac{1}{f^2}\del_X-\del_p\)s_\Omg+\frac{1}{2\mu}\(1-\frac{sT}{f^2}\del_X\)s_U=0 \label{s7}\\
&\frac{-2Tsc_n+2\mu sc_s-2\mu(nc_n+sc_s)+2w(nb_n+sb_s)}{ws(b_sc_n-b_nc_s)}+\(\del_\mu-\frac{1}{\mu}+\frac{w}{\mu f^2}\del_X\)s_U\no
&+\(\frac{1}{f^2}\del_X-\del_p\)s_V-\(\frac{1}{\mu f^2}\del_X-\frac{1}{\mu}\del_p\)s_\Omg=0 \label{s8}\\
&\frac{wT(nb_n+sb_s)+\mu T(nc_n+sc_s)+sT(-Tc_n+\mu c_s)}{w(b_sc_n-b_nc_s)}+\(\frac{n}{2}+\frac{sT}{2}\del_\mu\)s_U+\(w\del_p+\mu\del_\mu\)s_K\no
&+\(\frac{1}{2}-\frac{sT}{2}\del_p\)s_V-\(\frac{n}{2}\del_p+\frac{1}{2}\del_\mu\)s_\Omg=0. \label{s9}
\end{align}
\end{subequations}

There are seven unknown functions to be solved for from the nine equations. 
It is convenient to eliminate $s_u$ from (\ref{s8}) and (\ref{s9}) to obtain:
\begin{align}\label{s10}
-2\(\frac{w}{f^2}\del_X+\mu\del_\mu\)s_K+\(\frac{n}{f^2}\del_X+\del_\mu\)s_\Omg-\(1-\frac{sT}{f^2}\del_X\)s_V=0.
\end{align}
This is an over-determined problem. In 
order to proceed, we first take a moment to prove a commutation relation
between the following differential operators $\frac{1}{f^2}\del_X-\del_p$,
$\frac{w}{f^2}\del_X+\mu\del_\mu$ and $1-\frac{sT}{f^2}\del_X$. Bear in mind
that we have chosen $p,\,\mu,\,X$ as the independent thermodynamical variables.
$w,\,s,\,T,\,f^2$ contain implicit dependences on them through 
the equation of state. The dependences are subject to the Maxwell relations.
Starting from $dT=\frac{dp-n d\mu+f^2dX}{s}$, we have
\begin{align}
-\del_p\(\frac{n}{s}\)=\del_\mu\(\frac{1}{s}\),\quad -\del_X\(\frac{n}{s}\)=\del_\mu\(\frac{f^2}{s}\), \quad \del_X\(\frac{1}{s}\)=\del_p\(\frac{f^2}{s}\).
\end{align}
Using the above relations, it is a short exercise to show the operators 
$\frac{1}{f^2}\del_X-\del_p$, $\frac{w}{f^2}\del_X+\mu\del_\mu$ and $1-\frac{sT}{f^2}\del_X$ commute with each other.

To solve the equations (\ref{s1})-(\ref{s10}), we apply 
$\frac{1}{f^2}\del_X-\del_p$ to (\ref{s3})-(\ref{s4}) to obtain:
\begin{align}
\(\frac{1}{f^2}\del_X-\del_p\)\(\frac{2}{\mu}-\frac{sT}{\mu f^2}\del_X\)s_\omg-2\(\frac{1}{f^2}\del_X-\del_p\)\(2-\frac{sT}{f^2}\del_X\)s_B+\frac{2C\mu^2}{T}=0.
\end{align}
Using (\ref{s1}), (\ref{s2}) and noting
the commutation relation of the operators, we obtain
\begin{align}\label{sA}
\(\frac{1}{f^2}\del_X-\del_p\)s_K=0.
\end{align}
Note that $\frac{\del T}{\del X}=f^2\frac{\del T}{\del p}$, we conclude that
the solution is of the form $s_K=s_K(T,\mu)$. Plugging (\ref{s4}) and (\ref{s6}) into (\ref{s10}), we obtain:
\begin{align}
(1+\mu\del_\mu+\frac{\mu n}{f^2}\del_X)s_K=0.
\end{align}
With the specific form $s_K=s_K(T,\mu)$, we can solve the above by:
\begin{align}
s_K=\frac{f(T)}{\mu}.
\end{align}
The singular behavior of $s_K$ as $\mu\to 0$ forces $s_K$ to be zero.
To proceed further, we note from (\ref{s1}) and (\ref{s2}) that 
$-\frac{s_\omg}{\mu}+2s_B$ is also annihilated by $\frac{1}{f^2}\del_X-\del_p$.
Therefore, it can be parametrized as:
\begin{align}\label{sf}
-\frac{s_\omg}{\mu}+2s_B=-s_f(T,\mu).
\end{align}
Next, we plug (\ref{sf}) to (\ref{s3})-(\ref{s4}) and (\ref{s5})-(\ref{s6}),
the resultant PDEs give the following unique solution:
\begin{align}
s_f=\frac{2C\mu^2}{3T}.
\end{align}
It can be verified that the rest of the equations are identically satisfied.
We list our results on the coefficients of entropy current in the following:
\begin{align}\label{s_coeff}
&s_K=0 \no
&s_\omg=2\mu s_B-\frac{2C\mu^3}{3T} \no
&s_U=-2\(\frac{1}{f^2}\del_X-\del_p\)s_B \no
&s_V=-\frac{2C\mu}{T}+2\(\frac{n}{f^2}\del_X+\del_\mu\)s_B \no
&s_\Omg=-\frac{2C\mu^2}{T}+2\(1-\frac{sT}{f^2}\del_X\)s_B \no
&s_u=0.
\end{align}
Plugging (\ref{s_coeff}) back to (\ref{reduced}), we obtain the following 
nonvanishing non-entropy coefficients:
\begin{align}\label{ns_coeff}
&P_B=N_B=\Lam_B=0 \no
&\frac{l^2}{2}\Pi_L=\Pi_\omg,\quad V_L=\frac{sT}{w}\Pi_V,\quad \Pi_K=\frac{\Pi_V}{2},\quad \Pi_\Omg=\mu\Pi_V \no
&A_B=-\frac{C\mu^2}{w}-\frac{sT^2}{wf^2}\del_Xs_B \no
&\Lam_\omg+\frac{sT}{2w}\Pi_V=-\frac{2sT^2}{w(1+\mu b_X)}\bigg[\(\frac{f^2+c_X-nb_X}{sT}-(\del_X+c_X\del_p+b_X\del_\mu)\)\(\frac{sT}{f^2}\del_Xs_B\) \no
&+\frac{(f^2+c_X-\frac{w}{\mu}b_X)\mu^2}{sT^2}\bigg] \no
&N_\omg+\frac{\Pi_L}{2}=\frac{w}{sT}\(\Lam_\omg+\frac{sT}{2w}\Pi_V\) \no
&P_\omg=\((nc_n+sc_s)(-\frac{2}{s}+2T\del_p)+(nb_n+sb_s)(\frac{2n}{s}+2T\del_\mu)\)\(\frac{sT}{f^2}\del_Xs_B\)-\frac{w\mu}{sT}(nb_n+sb_s) \no
&\times\(\Lam_\omg+\frac{sT}{2w}\Pi_V\)-(nc_n+sc_s)\frac{2C\mu^2}{sT}+(nb_n+sb_s)\frac{2C\mu w}{sT} \no
&A_\omg=\(\frac{2(nb_n-c_n)}{s}-\frac{4\mu T}{w}+2T(c_n\del_p+b_n\del_\mu)\)\(\frac{sT}{f^2}\del_Xs_B\)-\frac{8C\mu^3}{3w}+(\frac{w}{\mu}b_n-c_n)\frac{2C\mu^2}{Ts} \no
&-\frac{\mu}{sT}(l^2+wb_n)\(\Lam_\omg+\frac{sT}{2w}\Pi_V\) \no
&V_\omg-\frac{l^2sT}{2w\mu}\Pi_\omg=-2T(1-\frac{2sT}{w})\frac{sT}{f^2}\del_Xs_B+\frac{8C\mu^2sT}{3w}-\frac{2C\mu^2}{3} \no
&V_B=\frac{sT^2}{w}\(\frac{C\mu}{T}-\frac{n}{f^2}\del_Xs_B\) \no
&V_U=\frac{2nT}{w}\(\frac{1}{f^2}\del_X-\del_p\)\(\frac{sT}{f^2}\del_Xs_B\) \no
&\mu V_K-\frac{V_\Omg}{2}=\frac{Cn\mu^2}{w}+\frac{nsT^2}{w}\(\frac{1}{sT}-\frac{1}{f^2}\del_X\)\(\frac{sT}{f^2}\del_Xs_B\) \no
&\mu V_V-V_\Omg=2nT\(\frac{1}{w}-\frac{1}{f^2}\del_X-\frac{\mu}{w}\del_\mu\)\(\frac{sT}{f^2}\del_Xs_B\)
\end{align}
Summarizing the results, we have determined, within the parity odd sector,
 the first order gradient corrections
to the stress tensor, charge current and ``Josephson'' equation up to five arbitray functions. The five functions are chosen as $s_B$, $\Pi_V$, $\Pi_\omg$, $V_\Omg$ and $T_\sig$. In particular, the entropy current, as well as the 
``Josephson'' equation is parametrized by the function $s_B$, and the rest 
enters the stress tensor and charge current. It is interesting to note
that terms proportional to the magnetic field is absent in the correction to
stress tensor, i.e. $P_B=N_B=\Lam_B=0$.
The presence of arbitrary functions in first order corrections is in contrast to
the case of normal hydrodynamics\cite{SS,oz}, where the positivity of entropy
production almost fully constrain the first order gradient corrections. We can
also compare our results with \cite{BBMY}(see also a more recent generalization in \cite{NO}). The authors of \cite{BBMY} studied the constraint of the positivity of entropy production in first order superfluid hydrodynamics, with both parity even and odd terms. A total of $6$ arbitrary functions are needed to parametrize the full gradient correction. $2$ of them appear in the entropy current. The authors claimed a nontrivial mixing between the parity odd and even sectors. While this is true in general, the nature of anomaly that is temperature and density independent(see for example \cite{Qian}), seems to suggest its induced effect to be non-dissipative(care is needed in a precise definition of anomaly induced terms). More recently the vanishing of the entropy production in the parity odd sector in the context of normal hydrodynamics has been argued in \cite{HUY} based on the principle of time reversal invariance. It will be interesting to see whether it is also true in the case of superfluid hydrodynamics.

\section{Conclusion}

We have extended the study of anomaly effect on
the constitutive equations of normal hydrodynamics\cite{SS,oz} 
to superfluid hydrodynamics. We have enumerated all possible $24$ parity odd,
first order gradient corrections to the constitutive equations:
$13$ for the stress tensor, $9$ for the charge current 
and $2$ for the ``Josephson''
equation. We used the existence of entropy current with vanishing 
entropy production to determine the transport coefficients. We found all
coefficients are uniquely determined up to $5$ arbitrary functions, with the
explicit expressions in (\ref{s_coeff}) and (\ref{ns_coeff}). We also found that the stress tensor does not receive correction from terms proportional to the magnetic field.

It will be interesting to extend the present analysis to chiral superfluid\cite{nuclear},
which is a $SU(N_f)\times SU(N_f)\rightarrow SU(N_f)$ superfluid. Rich
phenomena associated with anomaly effect on low energy finite density QCD
have been discovered over the past few years, see e.g.\cite{steph,spiral,YZ}. 
It is certainly desirable to formulate a superfluid hydrodynamical theory, 
which will be
useful for the exploration for the dynamical aspects of low energy QCD.

\noindent{\large \bf Acknowledgments} \vskip .35cm 

We would like to thank M. Ammon, J. Erdmenger, A. Rebhan, D. T. Son, A. O'Bannon
H. U. Yee and especially Ingo Kirsch for helpful discussions. We also thank
I. Kirsch and M. Lublinsky for useful comments and Shiraz Minwalla for pointing out an error in the previous version of the paper.
This work is supported by Alexander von Humboldt Foundation.

\appendix

\section{Derivations of $\del_\mu O^\mu$}

In this appendix, we want to express the divergence 
$\del_\mu O^\mu$($O=\omg,B,\Omg,K,U,V,L$) in terms of our basis set:
pseudovector$\cdot$vector and pseudoscalar$\cdot$scalar, by using the
ideal hydrodynamical equations. 

In $3+1$ dimensions, it is convenient to use vector notations. We will work
in the local rest frame(LRF) of the normal component, where $u^\mu=(1,0)$.
Note the condition $u^2=-1$ gives $\del_\mu u^0=0$, but $\del_\mu\vu\ne 0$.
We start by solving the following ideal hydrodynamical equations:

\begin{align}\label{ideal_hydro}
\left\{\begin{array}{l}
\del_\mu T^{\mu\nu}=F^{\nu\lam}j_\lam \\
\del_\mu j^\mu=0 \\
\mu=-\xi\cdot u \\
\del_\mu \xi_\nu-\del_\nu \xi_\mu=-F_{\mu\nu}
\end{array}
\right.,
\end{align}

Expressed in the vector notation, (\ref{ideal_hydro}) takes the following form:

\begin{align}\label{vec}
\left\{\begin{array}{l}
s\theta+\dot{s}=0 \\
w\dvu+\nabla p+f^2\nabla X-\vp\,\zeta=n\ve \\
\zeta+f^2\(\dot{\xi^0}+\nabla\cdot\vp\)+\del_tf^2\xi^0+\nabla f^2\cdot\vp=0 \\
\nabla\mu=\nabla \xi^0-\pdu \\
\dvp+\nabla \xi^0=\ve\\
\nabla\crs\vp=-\vb
\end{array}
\right.,
\end{align}

where the first equation is the conservation of entropy, which follows from
a combination of energy conservation and current conservation. The second and
the third equations are momentum and current conservation respectively. 
The fourth equation is from the ``Josephson'' equation.
The last two equations are vector form of the last equation of 
(\ref{ideal_hydro}). $\dvu$ is readily solved as 
$\dvu=\frac{1}{w}(n\ve-\nabla p-f^2\nabla X+\vp\,\zeta)$, which will be
repeatedly used below.

In LRF, the gradient terms $O^\mu$ has the following simple expression:

\begin{align}
&\vec{\omg}=\frac{1}{2}\nabla\crs\vu,\quad\vb=-\nabla\crs\vp \no
&\vec{\Omg}=\frac{1}{2}\vp\crs\dvu,\quad\vec{K}=-\vp\crs\ve \no
&\vec{U}=\frac{1}{2}\vp\crs\nabla p,\quad \vec{V}=\frac{1}{2}\vp\crs\nabla\mu
,\quad \vec{L}=\frac{1}{2}\vp\crs\pdu,
\end{align}

where the quantities with over left arrows $\pdu$ are to be contracted.
The divergences of $O^\mu$ are worked out in the following:

\begin{align}
\del_\mu\omg^\mu&=\frac{1}{2}\eps^{\mu\nu\alpha\beta}\del_\mu u_\nu\del_\alpha u_\beta=2\dvu\cdot\vec{\omg} \no
&=\frac{2}{w}(nE\cdot\omg-\omg^\mu\del_\mu p-f^2\omg^\mu\del_\mu X+\xi\cdot\omg\zeta)\\
\del_\mu B^\mu&=\frac{1}{2}\eps^{\mu\nu\alpha\beta}\del_\mu u_\nu F_{\alpha\beta}
=\dvu\cdot\vb-2\ve\cdot\vec{\omg} \no
&=\frac{1}{w}(nE\cdot B-B^\mu\del_\mu p-f^2B^\mu\del_\mu X+\xi\cdot B\zeta)-2E\cdot\omg\\
\del_\mu\Omg^\mu&=\frac{1}{2}\del_\mu P^{\mu\sig}\eps_{\sig\nu\alpha\beta}\xi^\nu
\del^\alpha u^\beta+\frac{1}{2}P^{\mu\sig}\eps_{\sig\nu\alpha\beta}\del_\mu \xi^\nu\del^\alpha u^\beta+\frac{1}{2}P^{\mu\sig}\eps_{\sig\nu\alpha\beta}\xi^\nu\del_\mu\del^\alpha u^\beta-\omg^\mu\del_\mu\mu-\mu\del_\mu\omg^\mu \no
&=-\theta \xi\cdot\omg+\xi^0\dvu\cdot\vec{\omg}-\frac{1}{2}\dvu\cdot\vb-\vec{\omg}\nabla\mu-\mu\del\cdot\omg+\frac{1}{2}\eps^{ijk}\xi_j\del_t\del_iu_k \no
&=-\theta \xi\cdot\omg-\frac{\mu}{2}\del\cdot\omg+\vec{\omg}\cdot\pdu
-\frac{1}{2}\del\cdot B-\ve\cdot\vec{\omg}+\frac{1}{2}\eps^{ijk}\xi_j\del_t\del_iu_k \no
&=-\theta \xi\cdot\omg-\frac{1}{2w}(-B^\mu\del_\mu p-f^2B^\mu\del_\mu X+nE\cdot B+\xi\cdot B\zeta)-\frac{\mu}{w}(nE\cdot\omg+\xi\cdot\omg\zeta\no
&-\omg^\mu\del_\mu p-f^2\omg^\mu\del_\mu X)+\omg^\mu \xi^\nu\del_\mu u_\nu+\frac{1}{2}\eps^{\mu\nu\alpha\beta}u_\nu\xi_\alpha u^\rho\del_\rho\del_\mu u_\beta \\
\del_\mu K^\mu&=\frac{1}{2}\del_\mu P^{\mu\sig}\eps_{\sig\nu\alpha\beta}\xi^\nu F^{\alpha\beta}+
\frac{1}{2}P^{\mu\sig}\eps_{\sig\nu\alpha\beta}\del_\mu \xi^\nu F^{\alpha\beta}-\del_\mu\mu B^\mu-\mu\del_\mu B^\mu \no
&=-\theta \xi\cdot B+\xi^0\dvu\cdot\vb-\dvu\cdot(\vp\crs\ve)+\frac{3}{2}\ve\cdot\vb-\vb\cdot\nabla\mu-\mu\del\cdot B+\frac{1}{2}\eps^{ijk}\xi_j\del_t\del_iA_k \no
&=-\theta \xi\cdot B+2\mu E\cdot\omg+\frac{1}{w}(-K^\mu\del_\mu p-f^2K^\mu\del_\mu X)+\frac{3}{2}E\cdot B-B^\mu\del_\mu \mu \no
&+\frac{1}{2}\eps^{\mu\nu\alpha\beta}u_\nu\xi_\alpha u^\rho\del_\rho\del_\mu A_\beta \\
\del_\mu U^\mu&=\frac{1}{2}\eps^{\mu\nu\alpha\beta}\del_\mu u_\nu \xi_\alpha\del_\beta p+\frac{1}{2}\eps^{\mu\nu\alpha\beta}u_\nu\del_\mu \xi_\alpha\del_\beta p \no
&=\frac{1}{2}\dvu\cdot(\vp\crs\nabla p)-\mu\vec{\omg}\nabla p-\xi\cdot\omg\dot{p}-\frac{1}{2}\vb\cdot\nabla p\no
&=\frac{1}{2w}(nK^\mu\del_\mu p-2f^2U^\mu\del_\mu X)-\mu\omg^\mu\del_\mu p-\xi\cdot\omg u^\nu\del_\nu p-\frac{1}{2}B^\mu\del_\mu p \\
\del_\mu V^\mu&=\frac{1}{2}\eps^{\mu\nu\alpha\beta}\del_\mu u_\nu \xi_\alpha\del_\beta\mu+\frac{1}{2}\eps^{\mu\nu\alpha\beta}u_\nu\del_\mu \xi_\alpha\del_\beta\mu \no
&=\frac{1}{2}\dvu\cdot(\vp\crs\nabla\mu)-\mu\vec{\omg}\cdot\nabla\mu-\xi\cdot\omg\dot{\mu}-\frac{1}{2}\vb\cdot\nabla\mu \no
&=\frac{1}{2w}(nK^\mu\del_\mu\mu+2U^\mu\del_\mu\mu-2f^2V^\mu\del_\mu X)-\mu\omg^\mu\del_\mu\mu-\xi\cdot\omg u^\mu\del_\mu\mu-\frac{1}{2}B^\mu\del_\mu\mu \no
\del_\mu L^\mu&=\frac{1}{2}\del_\mu \xi^\lam\eps^{\mu\nu\alpha\beta}u_\nu \xi_\alpha\del_\beta u_\lam+\frac{1}{2}\xi^\lam\eps^{\mu\nu\alpha\beta}\del_\mu u_\nu \xi_\alpha\del_\beta u_\lam+\frac{1}{2}\xi^\lam\eps^{\mu\nu\alpha\beta}u_\nu\del_\mu \xi_\alpha\del_\beta u_\lam \no
&=\frac{1}{2}\eps_{ijk}\xi_j\del_i\xi_l\del_k u_l+\frac{1}{2}\dvu\cdot(\vp\crs\pdu)
-\mu\vec{\omg}\pdu-\xi\cdot\omg\vp\cdot\dvu-\frac{1}{2}\vb\cdot\pdu \no
&=\frac{1}{2}\eps_{ijk}\xi_j\del_i\xi_l\del_k u_l+\frac{1}{w}(\frac{nK^\mu \xi^\nu\del_\mu u_\nu}{2}-L^\mu\del_\mu p-f^2L^\mu\del_\mu X)-\mu\omg^\mu \xi^\nu\del_\mu u_\nu \no
&-\frac{\xi\cdot\omg}{w}(nE\cdot \xi-l^\mu\del_\mu p-f^2l^\mu\del_\mu X+l^2\zeta)-\frac{1}{2}B^\mu \xi^\nu\del_\mu u_\nu.
\label{divL}
\end{align}
In (\ref{divL}) the first term is expressed as follows:

\begin{align}
\eps_{ijk}\xi_j\del_i\xi_l\del_k u_l&=\frac{1}{2}\sig\cdot\blam+\xi\cdot\omg\(\frac{\zeta+\mu\del_tf^2\vp\cdot\nabla f^2}{f^2}+\dot{f^2}+\frac{nE\cdot\xi-\vp\nabla p-f^2\vp\nabla X+l^2\zeta}{w}\) \no
&-\frac{1}{3}\xi\cdot B\theta+\vec{\omg}\cdot(\nabla X+\mu(\nabla\mu+\pdu))+\frac{1}{2}\sig_{\mu\nu}l^\mu B^\nu
\end{align}

\section{Relations among pseudovector$\cdot$vector}

The redundant pseudovector$\cdot$vector are related to those chosen in
(\ref{basis}) and pseudoscalar$\cdot$scalar as follows:
\begin{align}\label{pvvpss}
&\sig_{\mu\nu}l^\mu\omg=\omg^\mu\xi^\nu\del_\mu u_\nu-\xi\cdot\omg\frac{\theta}{3} \no
&\sig_{\mu\nu}l^\mu\Omg^\nu=-\frac{1}{2w}(nK^\mu \xi^\nu\del_\mu u_\nu-2L^\mu\del_\mu p-2f^2L^\mu\del_\mu X)+\frac{\xi\cdot\omg}{2w}(nE\cdot \xi-l^\mu\del_\mu p-f^2l^\mu\del_\mu X) \no
&-\frac{l^2}{2w}(nE\cdot\omg-\omg^\mu\del_\mu p-f^2\omg^\mu\del_\mu X) \\
&\sig_{\mu\nu}l^\mu K^\nu=K^\mu \xi^\nu\del_\mu u_\nu-E\cdot \xi\, \xi\cdot\omg+l^2E\cdot\omg\\
&\sig_{\mu\nu}l^\mu U^\nu=-\frac{1}{2}(2L^\mu\del_\mu p-\xi\cdot\omg l^\mu\del_\mu p+l^2\omg^\mu\del_\mu p)\label{lmu}\\
&\sig_{\mu\nu}l^\mu V^\nu=-\frac{1}{2}(2L^\mu\del_\mu\mu-\xi\cdot\omg l^\mu\del_\mu\mu+l^2\omg^\mu\del_\mu\mu)\\
&\sig_{\mu\nu}l^\mu L^\nu=\frac{1}{2}\xi\cdot\omg\sig_{\mu\nu}l^\mu l^\nu-\frac{l^2}{2}\sig_{\mu\nu}l^\mu\omg^\nu\\
&\Omg^\mu \xi^\nu\del_\mu u_\nu=-\frac{1}{w}(\frac{n}{2}K^\mu \xi^\nu\del_\mu u_\nu-L^\mu\del_\mu p-f^2L^\mu\del_\mu X)\\
&U^\mu \xi^\nu\del_\mu u_\nu=-L^\mu\del_\mu p\\
&V^\mu \xi^\nu\del_\mu u_\nu=-L^\mu\del_\mu\mu\\
&L^\mu \xi^\nu\del_\mu u_\nu=0\\
&\Omg^\mu\del_\mu p=-\frac{1}{w}(\frac{n}{2}K^\mu \xi^\nu\del_\mu u_\nu-f^2U^\mu\del_\mu X)\\
&U^\mu\del_\mu p=0 \\
&V^\mu\del_\mu p=-U^\mu\del_\mu \mu\\
&V^\mu\del_\mu\mu=0\\
&\Omg^\mu\del_\mu X=-\frac{1}{2w}(nK^\mu\del_\mu X+2U^\mu\del_\mu X)\\
&\Omg^\mu\del_\mu\mu=-\frac{1}{2w}(nK^\mu\del_\mu\mu+2U^\mu\del_\mu\mu-2f^2V^\mu\del_\mu X)
\end{align}
\begin{align}
&F_{\mu\nu}\xi^\nu \omg^\mu=\sig_{\mu\nu}l^\mu B^\nu+\xi\cdot B\frac{\theta}{3}-B^\mu\xi^\nu\del_\mu u_\nu\\
&F_{\mu\nu}\xi^\nu B^\mu=0\\
&F_{\mu\nu}\xi^\nu\Omg^\mu=-\frac{1}{2w}(nE_\mu-\del_\mu p-f^2\del_\mu X+\xi_\mu\zeta)(\xi\cdot B\xi^\mu-l^2B^\mu) \\
&F_{\mu\nu}\xi^\nu K^\mu=E\cdot\xi\xi\cdot B-l^2E\cdot B\\
&F_{\mu\nu}\xi^\nu U^\mu=-\frac{1}{2}l^\mu\del_\mu p\xi\cdot B+\frac{1}{2}l^2B^\mu\del_\mu p\\
&F_{\mu\nu}\xi^\nu V^\mu=-\frac{1}{2}l^\mu\del_\mu \mu\xi\cdot B+\frac{1}{2}l^2B^\mu\del_\mu \mu\\
&F_{\mu\nu}\xi^\nu L^\mu=-\frac{1}{2}l^2F_{\mu\nu}\xi^\nu \omg^\mu-\frac{1}{2}\xi\cdot B\sig_{\mu\nu}l^\mu l^\nu+\frac{1}{2}l^2\sig_{\mu\nu}l^\mu B^\nu\\
&E\cdot\Omg=-\frac{1}{2w}(K^\mu\del_\mu p+f^2K^\mu\del_\mu X)\\
&E\cdot K=0\\
&E\cdot U=\frac{1}{2}K^\mu\del_\mu p\\
&E\cdot V=\frac{1}{2}K^\mu\del_\mu\mu\\
&E\cdot L=\frac{1}{2}K^\mu \xi^\nu\del_\mu u_\nu \label{pvvpsse}.
\end{align}
Note that the scalar $l^\mu\del_\mu\mu$ in (\ref{lmu}) is not listed as a independent
scalar. By ideal hydrodynamical equations. it can be expressed as:
\begin{align}
&l^\mu\del_\mu\mu=u^\mu\del_\mu X-\mu u^\mu\del_\mu\mu+E\cdot \xi\frac{sT}{w}
-\frac{\mu l^2}{w}\zeta+\frac{\mu}{w}(l^\mu\del_\mu p+f^2l^\mu\del_\mu X)-\sig_{\mu\nu}l^\mu l^\nu-\frac{l^2}{3}\theta.
\end{align}

\vskip 1cm

\end{document}